\documentclass[%
 twocolumn,
 aps, prresearch,
 amsmath,amssymb,
 superscriptaddress
]{revtex4-1}

\usepackage{graphicx}
\usepackage{dcolumn}
\usepackage{bm}

\usepackage[utf8]{inputenc}
\usepackage[T1]{fontenc}
\usepackage{mathptmx}
\usepackage{graphicx}
\usepackage{dcolumn}
\usepackage{bm}
\usepackage{xcolor}
\definecolor{blue}{rgb}{0,0.5,1.0}
\usepackage[export]{adjustbox}
\usepackage{braket}
\usepackage{amsmath}
\usepackage{MnSymbol}
\usepackage{mathtools}

\begin{document}

\title{Transport through a quantum critical system: A thermodynamically consistent approach}

\author{C. W. W\"achtler}
\email{christopher.w.waechtler@campus.tu-berlin.de}
\affiliation{Institute of Theoretical Physics, Secr. EW 7-1, Technical University Berlin, Hardenbergstr. 36, D-10623 Berlin, Germany}
\author{G. Schaller}
\affiliation{Institute of Theoretical Physics, Secr. EW 7-1, Technical University Berlin, Hardenbergstr. 36, D-10623 Berlin, Germany}

\date{\today}

\begin{abstract}
Currents through quantum systems may probe non-analyticities in quantum-critical many-body ground states.
For a large class of dissipative quantum critical systems we show that it is possible to obtain the 
reduced system dynamics in the vicinity of quantum critical points in a thermodynamically consistent way, 
while capturing non-Markovian effects. We achieve this by combining reaction coordinate mappings with polaron transforms. 
Exemplarily, we consider the Lipkin-Meshkov-Glick model in a transport setup, where the
quantum phase transition manifests itself in the heat transfer statistics.
\end{abstract}

\maketitle

\section{\label{sec:Introduction} Introduction}

At vanishing temperature, a quantum many-body system may exhibit a drastic change upon modification of a control parameter solely driven by quantum fluctuations. 
Such a quantum phase transition (QPT) is accompanied by a closing gap of the low excitation energies and non-analytic changes of the ground state and observables~\cite{SachdevBook1999, EmaryBrandesPRE2003, LambergEtALPRL2004, RibeiroEtAlPRL2007, BastarracheaEtAlPRA2014}. 
Recent experiments have demonstrated the ability to investigate quantum phase transitions in ultracold atoms~\cite{BaumannEtAlNature2010, NagyEtAlPRL2010, BaumannEtAlPRL2011, HamnerEtAlNatureCom2014, TrenkwalderEtAlNaturePhys2016}, through cavity-assisted Raman transitions~\cite{BadenEtAlPRL2014}, in 1D ferromagnets~\cite{ColdeaEtAlScience2010}, spinor Bose-Einstein condensates~\cite{ZiboldEtAlPRL2010} or even by means of trapped ion quantum simulators~\cite{IslamEtAlNatCom2011, ZhangEtAlNature2017} and in circuit quantum electrodynamic lattices~\cite{FitzpatrickORX2017}. 
These engineered systems allow us to study a broad range of quantum critical phenomena in a highly controlled manner. 
However, as experimental setups are intrinsically open and often involve driven-dissipative systems~\cite{DomokosRitschPRL2002, BlackEtAlPRL2003, DimerEtAlPRA2007, NagyEtAlPRL2010, BaumannEtAlNature2010, BaumannEtAlPRL2011, BrenneckeEtAlPNAS2013, ZhiqiangEtAlOptica2017} that cannot be described by equilibrium models~\cite{KirtonEtAlAQT2019}, exploring the influence of nonequilibrium environments on QPTs and many-body physics is essential.
Examples include periodically driven systems~\cite{BastidasEtAlPRL2012, BastidasEtAlPRA2012, EngelhardtEtAlPRE2013, BastidasEtAlPRA2014}, quenched systems~\cite{DziarmagaPRL2005, RossiniEtAlPRL2009, PolkovnikovEtAlRevModP2011, AcevedoEtAlNJP2015, CampbellPRB2016, KopylovEtAlPRE2017}, systems with dissipation~\cite{MostameEtAlPRA2010,KeelingEtAlPRL2010, NagyEtAlPRA2011, BhaseenEtAlPRA2012, OztopEtAlNJP2012, KopylovEtAlPRA2013, GrimsmoParkinsJoPB2013, DallaTorreEtAlPRA2013, GenwayEtAlPRL2014, KirtonKeelingPRL2017} or critical transport setups~\cite{VoglEtAlAoP2011, VoglEtAlPRL2012, SchallerEtAlJoP2014, NagyDomokosPRL2015}.

A natural question that arises is whether signatures of quantum criticality can be probed when the system is coupled to reservoirs in a transport setup, such that even at steady state energy is transferred between the reservoirs through the system.
To establish a consistent formalism for such a transport scenario, two fundamental constraints have to be considered. 
Firstly, in the thermodynamic limit, the vanishing energy scales of low energy excitations lead to a breakdown of the standard perturbative expansion in the system-bath coupling.
Secondly, the developed framework has to obey the laws of thermodynamics, in particular when considering critical systems as working fluid of heat engines~\cite{CampisiFazioNatureCom2016, kloc2019a}. 

In general, there has been a great effort in developing techniques to access the strong coupling regime with master equations, such as polaron transformations ~\cite{BRandesVorrathIJMPB2003, ThorwartEtAlCP2004, KirtonKeelingPRL2013, XuCaoFronPhys2016} or the reaction coordinate (RC) mapping~\cite{IlesSmithEtAlPRA2014, IlesSmithJCP2016, StrasbergEtAlNJP2016, NewmanEtAlPRE2017, RestrepoEtAlNJP2018, NazirSchallerBook2018}. 
While the first approach is capable of addressing quantum-critical systems~\cite{KopylovSchallerPRA2019}, its thermodynamic interpretation remains 
challenging as system and reservoirs are transformed globally and a clear separation is not evident. 
On the other hand, the RC mapping comes with well-defined thermodynamic notions~\cite{StrasbergEtAlNJP2016}. 
However, when combining it with a secular approximation to obtain a Lindblad master equation, the approach becomes questionable when 
the energy gaps of the (transformed) system are comparable to or smaller than the (transformed) system-reservoir coupling strength.

\begin{figure}
\includegraphics[width=\columnwidth]{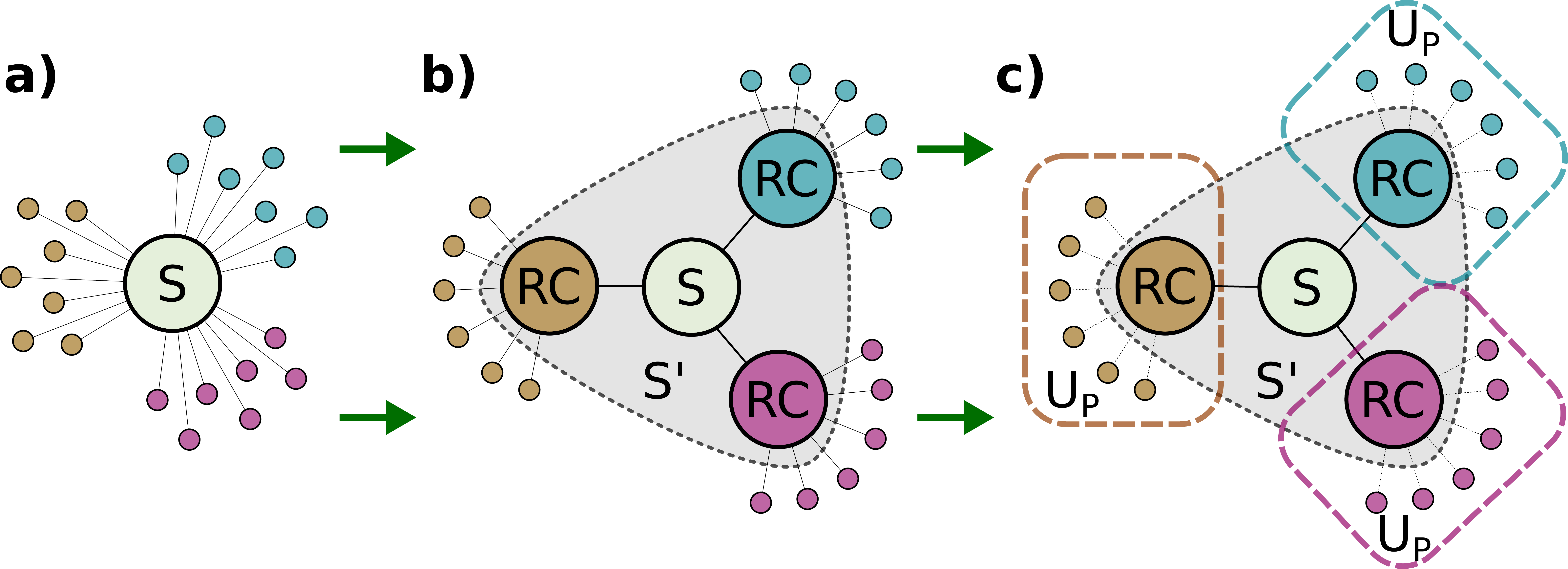}
\caption{Sketch of the method: 
a) A system $S$ is coupled to multiple heat reservoirs. 
b) After local RC mappings, the coupling to residual reservoirs is mediated by RCs. 
The new supersystem $S'$ (shaded region) consists of $S$ and the RCs. 
c) The polaron transform $U_\text{P}$, acting solely on original reservoir parts, alters the coupling between $S'$ and heat reservoirs while leaving $S$ unchanged, thereby allowing a weak coupling treatment across the full phase diagram.}
\label{fig:Transformations}
\end{figure}

In this work, we present a novel method to overcome this limitation.
It allows us to go beyond the perturbative weak coupling regime and describe quantum critical systems coupled to multiple structured heat baths while being thermodynamically consistent: 
Within the framework of the RC formalism, parts of the environment that interact strongly with the system can be defined as part of a supersystem, which in turn is coupled to effectively Markovian residual reservoirs (see Fig.~\ref{fig:Transformations}a and b). 
Applying a consecutive polaron transformation only on the original reservoir parts (see Fig.~\ref{fig:Transformations}c) allows for a perturbative treatment arbitrarily close to quantum-critical points. 
Moreover, the steady-state heat flow between supersystem and reservoirs is well defined and allows us to investigate the manifestation of QPTs in thermodynamic quantities.

\section{\label{sec:PerturbativeTreatment} Perturbative treatment of open quantum critical systems}
\subsection{\label{subsec:Model} Quantum critical system interacting with several heat baths}

We consider a class of systems, which undergo a QPT upon changing a control parameter $\kappa$ in the thermodynamic limit $N\to\infty$. 
After appropriate diagonalization, they can be described  by the Hamiltonian 
\begin{equation}
H_\text S = \sum_{n\ge 0} E_n(\kappa) \ket{n(\kappa)}\bra{n(\kappa)}, 
\end{equation}
where $E_n(\kappa)$ are the ordered energies and $\ket{n(\kappa)}$ the many-particle eigenstates of the system.
At least the lowest many-particle excitation energy $E_1(\kappa)-E_0(\kappa)$ vanishes at the critical point $\kappa_\text{cr}$, where the $\ket{n(\kappa)}$ undergo a non-analytic transition.
Examples of such systems described by $H_\text S$ are the Dicke model~\cite{DickePR1954, ArimondoBook1996, EmaryBrandesPRE2003, GarrawayPTRS2011} realized in Bose-Einstein condensates~\cite{BaumannEtAlNature2010, NagyEtAlPRL2010, BaumannEtAlPRL2011, HamnerEtAlNatureCom2014, BadenEtAlPRL2014, TrenkwalderEtAlNaturePhys2016}, the  Lipkin-Meshkov-Glick (LMG) model~\cite{MeshkovGlickLipkinNP1965, GilmorePLA1976, LeyvrazHeissPRL2005, SorokinEtAlPRE2014} or spinor Bose-Einstein condensates~\cite{ZiboldEtAlPRL2010, StamperKurnUedaRMP2013}, or the quantum Ising model~\cite{SachdevBook1999,dziarmaga2005a} and its quantum simulator realizations~\cite{friedenauer2008a,mostame2008a,zhang2009a,edwards2010a}.

We consider the scenario where the generic quantum-critical model is interacting with several bosonic heat reservoirs $\nu$ (see Fig.~\ref{fig:Transformations}a) 
$H_\text B^\nu = \sum_{k_\nu} \omega_{k_\nu} c^\dagger_{k_\nu} c_{k_\nu}$ with frequencies $\omega_{k_\nu}$ and bosonic annihilation operators $c_{k_\nu}$.
The heat baths are assumed at local equilibrium states with inverse temperatures $\beta^\nu = (k_\text B T^\nu)^{-1}$. 
To ensure that the system is thermodynamically stable~\cite{FordEtAlPRA1988}, it is required that the spectrum of the total Hamiltonian $H_\text{tot}$ is bounded from below for all values of the system-reservoir interaction strength.
This is manifest by writing the system-reservoir coupling via a generic dimensionless system operator $X_\nu=X_\nu^\dagger$ in terms of positive operators:
\begin{equation}
\label{eq:Htot1}
H_\text{tot} = H_\text S + \sum\limits_{\nu, k_\nu} \omega_{k_\nu}\left[c_{k_\nu}^\dagger + \frac{t_{k_\nu}}{\omega_{k_\nu}} X_\nu\right]\left[c_{k_\nu} + \frac{t_{k_\nu}}{\omega_{k_\nu}} X_\nu\right],
\end{equation} 
where $t_{k_\nu}\in\mathbb{R}$ represent emission (absorption) amplitudes that fix the spectral densities of the reservoirs $J_0^\nu(\omega) = 2\pi \sum_{k_\nu} t_{k_\nu}^2\delta(\omega-\omega_{k_\nu})$. 
In the standard weak-coupling approach (perturbative treatment of the $t_{k_\nu}$), the term quadratic in $X_\nu$ can be neglected, such that $X_\nu$ induces transitions between the unperturbed energy eigenstates of $H_\text S$ (Pauli master equation), leading to local thermalization in case of just one reservoir.
However, this naive perturbation theory will fail in the vicinity of the critical point, where the system-reservoir coupling strength exceeds (at least the smallest) system energy differences, manifest e.g. in second-order eigenvalue perturbation theory.
We argue that to maintain thermodynamic consistency, the quadratic term in $X_\nu$ should generally be kept in particular near critical points.

\subsection{\label{subsec:RCMapping} Reaction coordinate mapping and polaron transformation}

We propose to apply two consecutive transformations to each individual reservoir $\nu$ in order to apply weak-coupling theory while being thermodynamically consistent:

First, the RC mapping~\cite{IlesSmithEtAlPRA2014, IlesSmithJCP2016, StrasbergEtAlNJP2016, NewmanEtAlPRE2017, StrasbergEspositoPRE2017, SchallerEtAlPRB2018, StrasbergEtAlPRB2018}, which extracts a collective mode $b_\nu$ from the reservoir and introduces it as part of a new supersystem $S'$ (see Fig.~\ref{fig:Transformations}b):
\begin{equation}
\label{eq:TotalHamiltonian2}
H_\text{tot} = H_\text{S'} + \sum\limits_{\nu, k_\nu}\Omega_{k_\nu}\left[d_{k_\nu}^\dagger + \frac{h_{k_\nu}}{\Omega_{k_\nu}}(b_\nu^\dagger + b_\nu)\right]\left[d_{k_\nu} + \frac{h_{k_\nu}}{\Omega_{k_\nu}}(b_\nu^\dagger + b_\nu)\right],
\end{equation} 
where $S'$ is described by the Hamiltonian 
\begin{equation}
\label{eq:HSPrim2}
H_\text{S'} = H_\text S + \sum\limits_\nu \lambda_\nu\left[b^\dagger_\nu + \frac{g_\nu}{\lambda_\nu}X_\nu\right]\left[b_\nu + \frac{g_\nu}{\lambda_\nu}X_\nu\right]. 
\end{equation}
The RC mapping is a normal mode transformation of the original reservoir modes which is fully determined by the knowledge of $J_0^\nu(\omega)$ only~\cite{GargEtAJCP1985}. 
Thus, the RC frequencies $\lambda_\nu>0$, the coupling constants $g_\nu\in\mathbb{R}$ and the transformed residual spectral densities 
$J_1^\nu(\omega)=2\pi\sum_{k_\nu} h_{k_\nu}^2\delta(\omega-\Omega_{k_\nu})$, are fixed by the original spectral density $J_0^\nu(\omega)$ (see App.~\ref{app:RC}). 
We assume that the residual reservoirs are effectively Markovian, that is, the residual spectral densities are (super) ohmic and admit a perturbative treatment. 
If this is not the case, such mappings can be performed iteratively, which may result in a chain of RCs~\cite{MartinazzoEtAlJCP2011, StrasbergEtAlNJP2016} or more complicated geometries~\cite{huh2014a} until the resulting spectral densities are unstructured.
Still, the energy scales of $H_\text{S'}$ may become small at $\kappa_\text{cr}$ in comparison to any finite residual coupling.

Second, we therefore apply reservoir-specific polaron transformations $U_\text P^\nu = \exp\left[-(b_\nu^\dagger+b_\nu)P_\nu\right]$, 
where $P_\nu = \sum_{k_\nu}h_{k_\nu}/{\Omega_{k_\nu}}(d_{k_\nu}^\dagger - d_{k_\nu})$~\cite{BrandesPR2005, GlazmanShekhterSP1988, MahanBook2013, SchallerEtAlNJP2013, WingreenEtAlPRL1988}.
These commute mutually and also with $H_\text{S}$ (see Fig.\ref{fig:Transformations}c). 
Thereby, the original system remains unchanged and the total Hamiltonian (\ref{eq:TotalHamiltonian2}) takes with $U_\text{P}=\prod_\nu U_\text P^\nu$ in the polaron frame $H_\text{tot}' =U_\text{P}^\dagger H_\text{tot} U_\text P 
= U_\text P^\dagger H_\text{S'} U_\text P + \sum_{\nu,k_\nu} \Omega_{k_\nu} d_{k_\nu}^\dagger d_{k_\nu}$ the following form:
\begin{equation}
\label{eq:TotalHamiltonian3}
H_\text{tot}' = H_\text{S'}-\sum_\nu \lambda_\nu P_\nu^2 +\sum\limits_\nu \lambda_\nu\left(b_\nu - b_\nu^\dagger\right)P_\nu + \sum_{\nu,k_\nu} \Omega_{k_\nu}d_{k_\nu}^\dagger d_{k_\nu}, 
\end{equation}
where under the assumption that the residual reservoir coupling is weak $h_{k_\nu}/\Omega_{k_\nu}\ll 1$, we may also drop the quadratic term in $P_\nu$.
We observe that the residual reservoirs couple via their momenta to the RCs, which is inert to trivial displacements.
Furthermore, as the polaron transform is unitary, the energy scales of $U_\text P^\dagger H_\text{S'} U_\text P$ are just the same as that of $H_\text{S'}$, i.e., the effective coupling strength must scale adaptively with the phase parameter $\kappa$.
Therefore, we expect that when for $H_\text{tot}'$, Eq.~(\ref{eq:TotalHamiltonian3}), a second order perturbative treatment in $P_\nu$ is applicable away from the critical point, it will hold also for $\kappa \approx \kappa_\text{cr}$.

We stress the fact that a polaron transform without a prior RC mapping would have mixed system and reservoir observables, where a thermal state in the polaron frame would have a different interpretation in the original frame.
For the present approach, a perturbative treatment in $P_\nu$ yields 
$U_\text{P}^\dagger \exp\{-\beta_\nu \sum_{k_\nu} \omega_{k_\nu} c_{k_\nu}^\dagger c_{k_\nu}\} U_\text{P}\approx \exp\{-\beta_\nu \lambda_\nu b_\nu^\dagger b_\nu\} \exp\{-\beta_\nu  \sum_{k_\nu} \Omega_{k_\nu} d_{k_\nu}^\dagger d_{k_\nu}\}$, and for an ergodic evolution in the polaron frame, the standard thermodynamic consistency is expected.

\subsection{\label{subsec:Lindblad} Lindblad master equation}

To illustrate the framework introduced in Sec.~\ref{subsec:RCMapping}, we turn towards bosonizable systems for which the diagonalization of $H_\text{S'}$,  Eq.~(\ref{eq:HSPrim2}), can be performed explicitly, i.e., we consider systems with $N$ constituents that in the thermodynamic limit $N\to\infty$ can be approximately written as 
\begin{equation}
H_\text S = N E_\text G(\kappa) + \sum_{n\geq 0} \varepsilon_n(\kappa) a_n^\dagger a_n
\end{equation} 
with excitation energies $\varepsilon_n(\kappa)$ and bosonic modes $a_n$.
Assuming that these couple via their position to the reservoirs, we can insert the bosonization transformations for the coupling operator
\begin{equation}
X_\nu = \sum_{n\geq 0} C_{n\nu}(\kappa) \sqrt{N}+ D_{n\nu}(\kappa)\left(a_n^\dagger+a_n\right)/\sqrt{\varepsilon_n(\kappa)},
\end{equation}
where $C_{n\nu}(\kappa)$ and $D_{n\nu}(\kappa)$ are general functions. 
To account for a macroscopically populated ground state we introduce mean-fields $\alpha_n\in\mathbb{R}$ and $\gamma_\nu\in\mathbb{R}$ and new operators $A_n$ and $B_\nu$, such that $a_n = A_n + \sqrt{N}\alpha_n$ and $b_\nu = B_\nu + \sqrt{N}\gamma_\nu$, and decompose $H_\text{S'}$ in orders of $N^{-1/2}$, i.e. 
$H_\text{S'}=N H_0 + \sqrt{N}H_1 + H_2 + \mathcal O(N^{-1/2})$.
In order to expand around the correct ground state in the two phases (normal and symmetry broken) separated by $\kappa_\text{cr}$, one demands that $H_1$ is always equal to zero, which yields $\alpha_\nu =0$ and 
\begin{equation}
\gamma_\nu = \left\{\begin{array}{ll} 0 & \text{normal phase}\\
-\frac{g_\nu}{\lambda_\nu}\sum\limits_{n\geq 0}C_{n\nu}(\kappa) &\text{symmetry broken phase}\end{array}\right. .
\end{equation}
Then, $H_0 = E_\text G$ and the ground state energy remains unchanged. 

The quadratic Hamiltonian 
\begin{equation}
\begin{aligned}
H_2 &= \sum_{n\geq 0}\varepsilon_n a_n^\dagger a_n \\
&+ \sum_{n,\nu} \lambda_\nu \left[B_\nu^\dagger +\frac{g_\nu D_{n\nu}}{\lambda_\nu \sqrt{\varepsilon_n}}\left(a_n^\dagger + a_n\right)\right]\left[B_\nu +\frac{g_\nu D_{n\nu}}{\lambda_\nu \sqrt{\varepsilon_n}}\left(a_n^\dagger + a_n\right)\right]
\end{aligned}
\end{equation}
can be diagonalized by an orthogonal transformation $U$, such that $H_2 = \sum_{n\geq 0} \bar\varepsilon_n(\kappa) e^\dagger_n e_n$, where we have neglected the zero point energy. Hence, after diagonalization 
\begin{equation}
H_\text{S'} = NE_\text G(\kappa) +\sum_{n\geq 0} \bar \varepsilon_n(\kappa) e_n^\dagger e_n,
\end{equation} 
where $\bar \varepsilon_0(\kappa\to\kappa_\text{cr}) \to 0$. 
Note that the position of the QPT remains unchanged as the terms proportional to $N$ in $H_\text{S}$ and $H_\text{S'}$ are equal.
After applying the orthogonal (Bogoliubov) transformation $U$ that diagonalizes $H_\text{S'}$, to the system-reservoir coupling, the total Hamiltonian in the polaron frame, $H'_\text{tot}=U_\text P^\dagger H_\text{tot}U_\text P$, takes the simple form
\begin{equation}
\label{eq:TotalHamiltonain4}
H'_\text{tot} \approx H_{\text S'} - \sum\limits_{\nu,n} \left[U_n^{\nu} \sqrt{\bar\varepsilon_n \lambda_\nu} \left(e_n^\dagger -e_n\right)P_\nu + \sum\limits_k\Omega_{k_\nu}d_{k_\nu}^\dagger d_{k_\nu}\right], 
\end{equation}
where $U_n^\nu$ denote the entries of $U$ and we have neglected the term quadratic in $P_\nu$ (see Sec.~\ref{subsec:RCMapping}).
Collecting all factors in a polaron frame spectral density, we see that ${J'}_1^\nu(\omega) = (U_n^\nu)^2 \bar\varepsilon_n\lambda_\nu J_1^\nu(\omega)/\omega^2$.

As the interaction shows the same scaling behavior as the system, assuming that $\bar \varepsilon_n$ are non-degenerate away from the QPT, the Born-Markov secular approximations can be applied across the full phase diagram. 
Thus, the  reduced system density matrix $\varrho(t)$ evolves according to a Lindblad-type master equation, 
\begin{equation}
\begin{aligned}
\dot\varrho(t) &= -i[H_\text{S'},\varrho] +\sum\limits_\nu \mathcal L_\nu\varrho\\
&= -i[H_\text{S'},\varrho] +  \sum_{\nu,n}(F_n^\nu \mathcal D[e_n]\varrho + G_n^\nu\mathcal D[e_n^\dagger]\varrho)
\end{aligned}
\end{equation}
with transition rates
\begin{equation}
\label{eq:Rates}
\begin{aligned}
F_n^\nu &= \frac{{U_n^{\nu}}^2\lambda_\nu J_1^\nu(\bar \varepsilon_n)}{\bar \varepsilon_n}\left[n_\text B^\nu(\bar \varepsilon_n)+1\right],\\ 
G_n^\nu &= \frac{{U_n^{\nu}}^2 \lambda_\nu J_1^\nu(\bar\varepsilon_n)}{\bar\varepsilon_n}n_\text B^\nu(\bar \varepsilon_n).
\end{aligned}
\end{equation}
Here, $n_\text B^\nu(\omega) = [\exp(\beta^\nu \omega)-1]^{-1}$ and $\mathcal D[O]\varrho \equiv O\varrho O^\dagger - \frac{1}{2}\{O^\dagger O,\varrho\}$ for any operator $O$. 
We stress that the Markovian Lindblad equation for the supersystem captures non-Markovian effects in the original system.
In the long time limit $\varrho(t\to\infty) = \otimes_n \exp(-\bar \beta_n \bar\varepsilon_n e_n^\dagger e_n)/Z_n$ with individual partition functions 
$Z_n = \text{Tr}\{\exp(-\bar \beta_n \bar\varepsilon_n e_n^\dagger e_n)\}$, where the effective inverse temperature $\bar \beta_n$ is related to the emission and absorption rates by $\bar \beta_n \bar \varepsilon_n = -\ln(\sum_\nu G_n^\nu/\sum_\nu F_n^\nu)$. 

\subsection{\label{subsec:Heat} Heat transfer statistics and the second law}

As the local detailed balance condition $F_n^\mu G_n^\nu/G_n^\mu F_n^\nu = \exp[-\bar \varepsilon_n(\beta^\nu -\beta^\mu)]$ is fulfilled [see Eq.~(\ref{eq:Rates})], a transparent thermodynamic interpretation is possible. 
Based on the rigorous framework of full counting statistics~\cite{EspositoHarbolaMukamelRMP2009} and large deviation theory~\cite{VaradhanCPApplMAth1966, GartnerTPA1977, EllisAnnProb1984, GarrahanLesanovskyPRL2010, PigeonEtAlPRA2015}, we obtain the counting variable $\chi_n^\mu$ dependent cumulant generating function of the heat flow statistics in the long time limit ($t\to\infty$) of the exchanged energy between a reference reservoir $\mu$ and the supersystem $S'$ (see App.~\ref{app:FullCounting}), 
\begin{equation}
\label{eq:CumulantGenerating}
\mathcal C^\infty_\mu = \sum\limits_n \left\{\Delta^-_n - \sqrt{\left[\Delta_n^- + \frac{f_\mu^-(\chi_n^\mu)}{2}\right]^2- f_\mu^+(\chi_n^\mu)\left[\Delta_n^+ + \frac{f_\mu^+ (\chi_n^\mu)}{4}\right]}\right\}.
\end{equation}
Here, $\Delta^\pm_j \equiv \sum_\nu (F_n^\nu \pm G_n^\nu)/2$ and $f_\mu^\pm(\chi_n^\mu) [F_n^\mu (e^{\text{i}\chi_n^\mu \bar \varepsilon_n}-1)\pm G_n^\mu(e^{-\text{i}\chi_n^\mu \bar \varepsilon_n}-1)]$. 
The cumulant of order $k$ associated with the heat flow probability distribution is expressed in terms of derivatives of $\mathcal C^\infty_\mu$, Eq.~(\ref{eq:CumulantGenerating}), that is 
\begin{equation}
\llangle \dot Q^k\rrangle = \sum_n(-\text{i})^k \left.\frac{\partial^k \mathcal C^\infty_\mu}{\partial(\chi_n^\mu)^k}\right|_{\chi_n^\mu=0}.
\end{equation}
The additive decomposition of the generating function reflects the fact that in the diagonal frame, the bosonic modes act as independent transport channels generating independent stochastic events, which eventually renders all cumulants additive. 
Furthermore, $\mathcal C^\infty_\mu$ fulfills a Gallavotti-Cohen symmetry~\cite{LebowitzSpohnJSP1999, PigeonEtAlNJP2015}
with respect to $\chi_n^\mu \to -\text{i}(\beta^\mu-\sum_{\nu\neq \mu}\beta^\nu)-\chi_n^\mu$,
which is a direct consequence of the local detailed balance condition. 
Therefore, a steady state fluctuation theorem holds~\cite{EspositoHarbolaMukamelRMP2009}, which relates the probability $p(\{m_n\},t)$ that a net number of $m_n$ quanta haven been transferred along the $n$th-channel between the reference reservoir $\mu$ to the supersystem $S'$ in a time $t$, i.e., 
\begin{equation}
\lim_{t\to\infty}\frac{p(\{m_n\},t)}{p(\{-m_n\},t)} = \exp[(\sum_{\nu\neq \mu}\beta^\nu -\beta^\mu)\sum_n \bar \varepsilon_n m_n].
\end{equation} 
It follows from the fluctuation theorem that at quantum critical points the (net) heat transfer is blocked through the critical channel with $\bar \varepsilon_n\to 0$.

From the existence of a fluctuation theorem or via the use of Spohn's inequality one 
can show the non-negativity of the entropy production rate in a straightforward calculation: We introduce $\varrho_\text{eq}^\nu \equiv e^{-\beta^\nu H_\text{S'}}/Z_\nu$ for which $\mathcal L_\nu \varrho_\text{eq}^\nu=0$. 
The time derivative of the von-Neumann entropy is given by
\begin{equation}
\frac{d}{dt} S = -\text{Tr}\left\{\dot \varrho \ln \varrho\right\} = -\sum\limits_\nu \text{Tr}\left\{\mathcal L_\nu \varrho \ln \varrho\right\}.
\end{equation}
By use of Spohn's inequality for each individual reservoir, 
\begin{equation}
-\text{Tr}\left\{\mathcal L_\nu \varrho \left[\ln \varrho - \ln {\varrho^\nu_{\rm eq}}\right]\right\} \geq 0,
\end{equation}
and the definition of heat flow coming from reservoir $\nu$, $\left<\dot Q^\nu\right> = \text{Tr}\left\{H_\text{S'}\mathcal L_\nu \varrho\right\}$, it can be shown that the second law holds, i.e., that the entropy production rate is non-negative:
\begin{equation}
\dot S_\text{i} = \frac{d}{dt}S - \sum\limits_\nu \beta^\nu \left<\dot Q^\nu \right> \geq 0.
\end{equation}
This demonstrates the thermodynamic consistency of our approach. 
Moreover, the change of energy in the original reservoir $\nu$, $\langle\dot H_{\text{B}}^\nu\rangle $, is connected to the change in energy in the residual reservoir $\langle\dot H_\text{B'}^\nu\rangle $ through the energy change of the RC, i.e., $\langle\dot H_\text B^\nu\rangle  \approx \langle\dot H_\text{B'}^\nu +\dot H_\text{RC}^\nu\rangle $.
At steady state $\langle\dot H_\text{RC}^\nu\rangle  =0$, such that $\langle\dot H_\text B^\nu\rangle  \approx \langle U_\text P^\dagger \dot H_\text{B'}^\nu U_\text P\rangle $. 
We stress that without the RC mapping prior to the polaron transformation, system and reservoir would be mixed without a clear thermodynamic interpretation in contrast to the approach shown here. 

\section{\label{sec:LMG} Transport through the Lipkin-Meshkov-Glick model}

\begin{figure}
\includegraphics[width=\columnwidth]{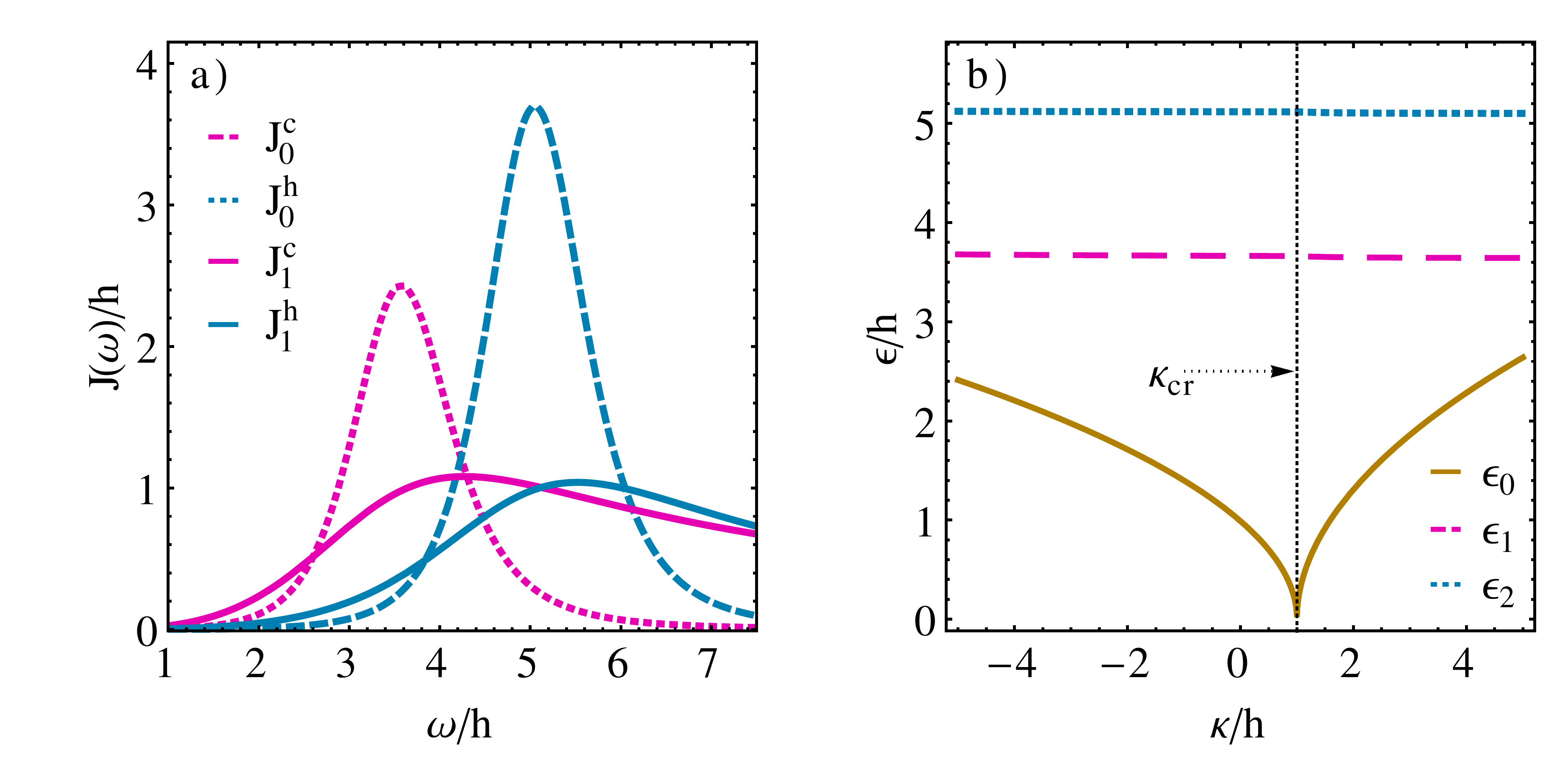}
\caption{a) Original spectral densities $J_0^\nu$ (dashed) of the hot ($\nu=\text h$) and cold ($\nu=\text c$) reservoir and $J_1^\nu$ after the RC mapping (solid) [see Eq.~\ref{eq:SpectralDensities}]  b) Excitation spectrum of the supersystem $S'$ after the RC mapping consisting of the LMG model and two RCs. At the critical parameter $\kappa_\text{cr}$ of the uncoupled LMG model the gap closes above the ground state marking the QPT. Parameters: $\Gamma_\text h = 300.0h$, $\Gamma_\text c = 140.0h$, $\delta_\text h = \delta_\text c= h$, $\bar\omega_\text c= 3.5h$ and $\bar\omega_\text h = 5.0h$.
}
\label{fig:LMG1}
\end{figure}

As a specific application of the general theory for bosonizable systems, we investigate the LMG model~\cite{MeshkovGlickLipkinNP1965, GilmorePLA1976, LeyvrazHeissPRL2005, SorokinEtAlPRE2014}, which describes $N$ two-level systems collectively interacting with an external field and among themselves, coupled to two reservoirs at different temperatures (hot and cold). 
In terms of collective spin operators $J_m = \sum_{n=1}^N \sigma_m^{(n)}/2$, where $m\in\{\text x,\text y,\text z\}$ and $J_\pm = J_\text{x}\pm \text{i} \cdot J_\text{y}$ with $\sigma_m^{(n)}$ denoting the Pauli matrix of the $n$th spin, the LMG Hamiltonian is given by
\begin{equation}
\label{eq:HLMG}
H_\text{S} = -h J_\text{z} - \frac{\kappa}{N}J_\text{x}^2\,,
\end{equation}
where $h$ is the strength of the magnetic field in z-direction and $\kappa$ denotes the coupling between the two-level systems. The scaling of $J_\text x$ with $1/\sqrt{N}$ ensures a meaningful thermodynamic limit ($N\to\infty$). The system undergoes a QPT at $\kappa_\text{cr} = h$ with non-analytic ground-state energy density~\cite{GilmorePLA1976, DusuelVidalPRL2004, VidalEtAlPRA2004, LeyvrazHeissPRL2005, DusuelVidalPRB2005, SorokinEtAlPRE2014}: For $\kappa<h$ (normal phase) the system has a unique ground state, whereas for $\kappa>h$ the system exhibits a \emph{symmetry-broken phase}~\cite{RibeiroEtAlPRL2007, HuangEtAlPRA2018} with e.g. collective spontaneous polarization and bifurcation of the $J_\text{z}$-expectation value. In the thermodynamic limit $H_\text{S}$ can be diagonalized by a Holstein-Primakoff transformation~\cite{HolsteinPrimakoffPR1940, KopylovEtAlPRA2013} and subsequent displacement of the bosonic operators~\cite{GlauberPR1963}, yielding $H_\text{S}=N E_\text G+ \varepsilon  a^\dagger a$~\cite{SorokinEtAlPRE2014, kopylov2019a}. 

\begin{figure}[]
\includegraphics[width=\columnwidth]{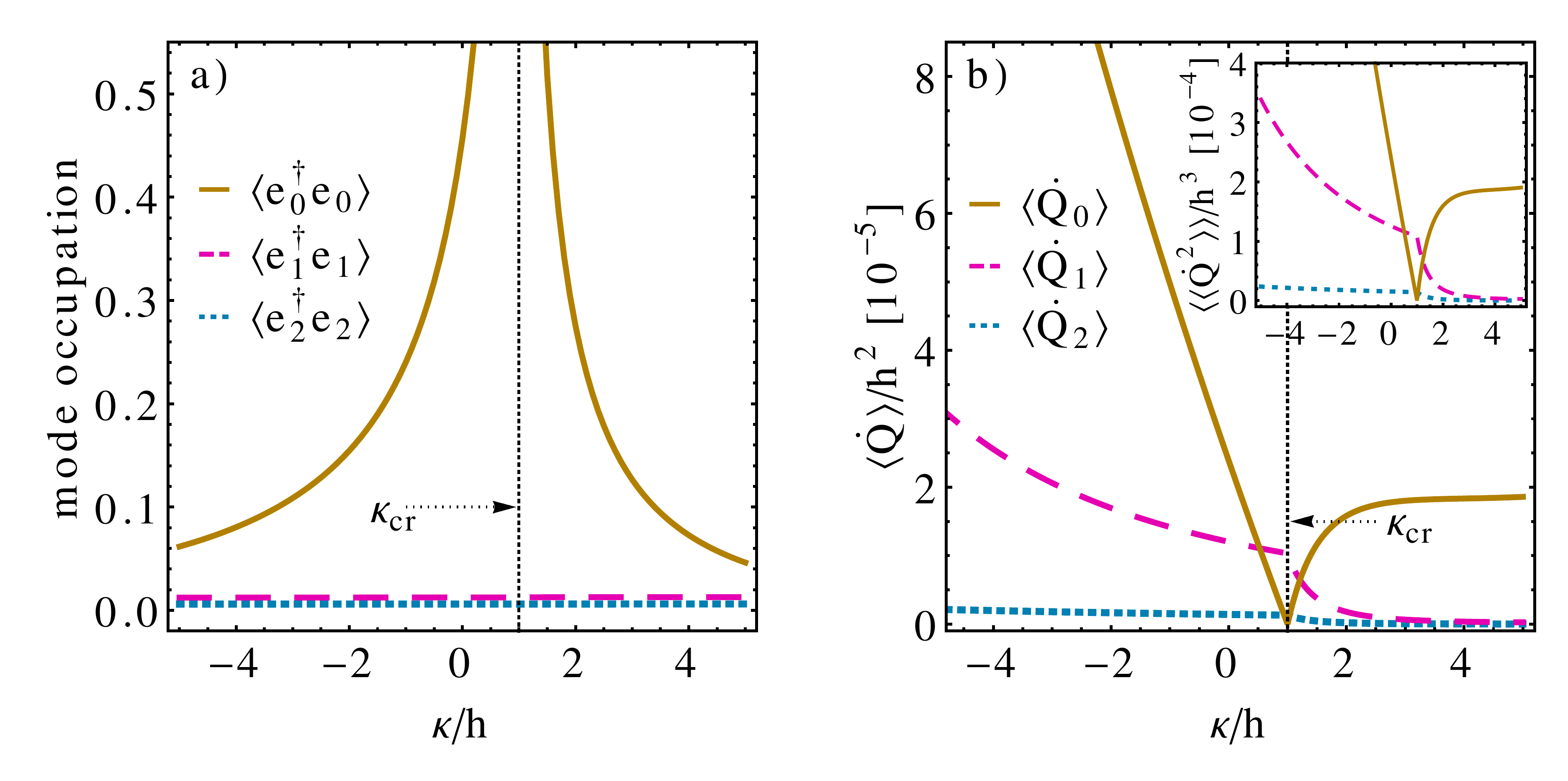}
\caption{a) Steady state occupation of the excitation modes for $\beta^\text h = 1.0 / h$ and $\beta^\text c = 1.2/h$. The diverging mode occupation $\langle e_0^\dagger e_0\rangle $ indicates the QPT. b) Steady state heat flow: At the QPT the heat flow $\langle \dot Q_0\rangle $ with vanishing excitation energy is blocked and in the symmetry-broken phase the total heat flow is significantly reduced. Inset: Second cummulant $\llangle \dot Q^2\rrangle$ of the heat flow statistics showing the same behavior as the average heat flow, especially non-analytic behavior at $\kappa = \kappa_{\text{cr}}$. Other parameters as in Fig. 2.
}
\label{fig:LMG2}
\end{figure}

The total system including the two reservoirs, hot ($\nu = \text h$) and cold ($\nu = \text c$), is described by Eq.~(\ref{eq:Htot1}) with $X_\nu = J_x/\sqrt{N}$. Choosing peaked original spectral densities of the reservoirs (see inset of Fig.~\ref{fig:LMG1} a), 
\begin{equation}
\label{eq:SpectralDensities}
J_0^\nu(\omega) = \Gamma_\nu \frac{\omega^3 \delta_\nu^5}{\left[\left(\omega-\bar\omega_\nu\right)^2+\delta^2\right]^2\left[\left(\omega+\bar\omega_\nu\right)^2+\delta^2\right]^2}
\end{equation}
results in unstructured spectral densities of the residual reservoirs $J_1^\nu(\omega)$ (see Fig.~\ref{fig:LMG1}a) after the RC mapping (see App.~\ref{app:RC}).
The supersystem $S'$ consisting of the LMG and two RCs [see Eq.~(\ref{eq:HSPrim2})] reads in the diagonal frame 
$H_\text{S'} = N E_\text G + \sum_{n=0}^2 \bar\varepsilon_n e_n^\dagger e_n$, where only $\bar\varepsilon_0(\kappa\to h)\to 0$ (see Fig.~\ref{fig:LMG1}b).
Following the treatment we present in this work, the steady state dynamics of the nonequilibrium LMG model are calculated straightforwardly. 
Before investigating the transport properties across the QPT, we analyze the system properties. To this end we look at the mean populations of the independent channels $\langle e_n^\dagger e_n\rangle  = \partial \ln Z/\partial(-\bar\beta_n \bar \varepsilon_n)$, which are shown in Fig.~\ref{fig:LMG2} a). The diverging occupation of the mode with vanishing excitation energy $\langle e_0^\dagger e_0\rangle $ indicates the QPT. However, the two additional modes of the supersystem, $\langle e_{1}^\dagger e_{1}\rangle $ and $\langle e_{2}^\dagger e_{2}\rangle$, are mostly effectively unoccupied, which shows that close to the quantum critical point, the low-temperature physics of the system is dominated by criticality. 

As system observables are often difficult to measure in an experiment, we also look for signatures of the QPT in the heat flows and the statistics thereof. The first cumulant with $k=1$ represents the average heat flow from the hot reservoir into the system $\llangle \dot Q\rrangle=\langle\dot Q\rangle=\sum_n \langle \dot Q_n\rangle $. Here, positive values of $\langle \dot Q\rangle$ indicate energy transfers from the hot bath into the system and vice versa. Another interesting quantity is the second cummulant $\llangle \dot Q^2_n\rrangle = \langle \dot Q_n^2\rangle - \langle \dot Q_n\rangle^2$ measuring the noise of each channel $n$. We show both of these quantities in Fig.~\ref{fig:LMG2} b and its inset, respectively. 
At the quantum critical point, the heat transfer along the transport channel with closing energy gap $\langle \dot Q_0\rangle$ vanishes as already indicated by the steady state fluctuation theorem.
Since this channel dominates the total heat flow, also the latter is significantly reduced at the critical point.
Moreover, in the symmetry broken phase $\kappa >\kappa_\text{cr}$ the ground state is macroscopically occupied, which suppresses the energy exchange along the system compared to the normal phase $\kappa <\kappa_\text{cr}$. Furthermore, the second cumulant scales equally as the first cumulant and also vanishes at the QPT. This behavior can be observed for all orders of the cumulants (not shown here), i.e., all cumulants of the heat flow statistics vanish at $\kappa_\text{cr}$. 

\section{\label{sec:Conclusions} Summary and conclusions}

In this work we have presented a general method to study nonequilibrium QPTs which is consistent with the laws of thermodynamics based on a combination of the RC mapping and a polaron transformation. This allows us to write the reduced system dynamics by means of a Lindblad type master equation arbitrarily close to quantum critical points and comes with a clear thermodynamic interpretation. For the specific example of the LMG model interacting with a hot and cold thermal reservoir we investigate the cumulants of the heat transfer statistics, which reflect the QPTs by non-analytic behavior at the critical point. 

We would like to remark, that if one would not write the total Hamiltonian in terms of positive operators [see Eq.~(\ref{eq:Htot1})] but neglect the squared term of $X_\nu$, the interaction with the reservoirs would shift the position of the QPT and, moreover, induce additional phase transitions. However, as shown here these additional phase transitions are prohibited in the same way as the diamagnetic term prevents the QPT of the Dicke model~\cite{RzazewskiPRL1975, RzazewskiPLA1976, RzazweskiPRA1976, KnightEtAlPRA1978, VukicsEtAlPRA2015, BialynickiRzazweskiPRA1979, SlyusarevYankelevichTMP1979}. 

Beyond generic open systems with small or vanishing energy gaps, the formalism presented here is particularly relevant for quantum critical systems like quantum Ising chains~\cite{VoglEtAlPRL2012}, cold atoms~\cite{ChienEtAlNaturePhys2015} and spinor Bose-Einstein condensates~\cite{ZiboldEtAlPRL2010}.
Exploring these critical systems is a natural next step and with the advances in quantum simulation, structured reservoirs~\cite{ChenEtAlPRL2017} as well as critical systems~\cite{IslamEtAlNatCom2011}, appropriate setups can be engineered to test our predictions.
Moreover, our approach may be extended to systems undergoing topological phase transitions, which also exhibit an energy gap closing, like the Su-Schrieffer-Heeger model~\cite{SuSchriefferHeegerPRL1979, HeegerEtAlRMP1988, AsbothEtAlBook2016, BoehlingEtAlPRB2018}, giving rise to interesting physics to investigate. 

\acknowledgements{
We gratefully acknowledge financial support from the Deutsche Forschungsgemeinschaft (DFG, German Research Foundation) through project BR1528/8-2. Furthermore, we are thankful for stimulating discussions with A. Knorr, S. Restrepo, S. B\"ohling and especially V. M. Bastidas and W. J. Munro.
}

\appendix
\section{\label{app:RC} Reaction coordinate mapping}

The idea of the reaction coordinate (RC) mapping is to introduce a part of the reservoir as part of an enlarged supersystem. 
We follow here the procedure discussed in~\cite{IlesSmithEtAlPRA2014, IlesSmithJCP2016, StrasbergEtAlNJP2016, NewmanEtAlPRE2017, RestrepoEtAlNJP2018, NazirSchallerBook2018}. 
We postulate the equivalence (up to a possible shift) of the Hamiltonians defined in Eqs.~(1-3). The mapping shall then be achieved by means of a Bogoliubov transform
\begin{equation}
c_{k_\nu} = u_{k0}^\nu b_\nu +\sum\limits_{q\geq 1}u_{kq}^\nu d_{q_\nu} + v_{k0}^\nu b_\nu^\dagger + \sum\limits_{q\geq 1}v_{kq}^\nu d_{q_\nu}^\dagger
\end{equation} 
and similar for the creation operator $c_{k_\nu}^\dagger$. To maintain the bosonic character of the new modes, the coefficients $u_{kq}^\nu$ and $v_{kq}^\nu$ are chosen via
\begin{equation}
\begin{aligned}
u_{kq}^\nu &= \frac{1}{2}\left(\sqrt{\frac{\omega_{k_\nu}}{\Omega_{q_\nu}}} + \sqrt{\frac{\Omega_{q_\nu}}{\omega_{k_\nu}}}\right) \Lambda_{kq}^\nu,\\
v_{kq}^\nu &= \frac{1}{2}\left(\sqrt{\frac{\omega_{k_\nu}}{\Omega_{q_\nu}}} - \sqrt{\frac{\Omega_{q_\nu}}{\omega_{k_\nu}}}\right) \Lambda_{kq}^\nu,
\end{aligned}
\end{equation}
with the unknown orthogonal transformation $\Lambda^\nu$ obeying $\sum_q \Lambda^\nu_{kq}\Lambda^\nu_{k'q} = \delta_{kk'}$. Here, $q=0$ maps to the annihilation and creation operators of the RC. 

By inserting the transformation and comparing the terms, we find expressions for the energy and coupling strength of the RC, 
\begin{equation}
\lambda_\nu^2 = \Omega_{0_\nu}^2 = \frac{\int\limits_0^\infty \omega J_0^\nu(\omega)d\omega}{\int\limits_0^\infty \frac{J_0^\nu(\omega)}{\omega}d\omega},\quad g_\nu^2 = \frac{1}{2 \pi \lambda_\nu}\int\limits_0^\infty \omega J_0^\nu(\omega) d\omega.
\end{equation}
Additionally, the transformed spectral density can be obtained from the original spectral density by the following transformation:
\begin{equation}
J_1^\nu(\omega)= \frac{{4}g_\nu^2 J_0^\nu(\omega)}{\left[\frac{1}{\pi}\mathcal P \int\limits_{-\infty}^\infty d\omega'~\frac{J_0^\nu(\omega')}{\omega'-\omega}\right]^2+\left[J_0^\nu(\omega)\right]^2}.
\end{equation}
Here, $\mathcal P$ indicates the principal value and it is understood that $J_0^\nu(\omega)$ is extended to negative values of $\omega$ via $J_0^\nu(-\omega) =-J_0^\nu(\omega)$.

For the specific choice
\begin{equation}
J_0^\nu(\omega) = \Gamma_\nu \frac{\omega^3 \delta_\nu^5}{\left[\left(\omega-\bar\omega_\nu\right)^2+\delta^2_\nu\right]^2\left[\left(\omega+\bar\omega_\nu\right)^2+\delta^2_\nu\right]^2}
\end{equation}
of the original spectral density, the residual spectral density after the RC mapping can be calculated analytically
\begin{equation}
J_1^\nu(\omega) = \frac{16 \delta_\nu^3 \omega^3}{\sqrt{\delta_\nu^2+\bar\omega_\nu^2} \left[\left(\delta_\nu ^2+\bar\omega^2_\nu\right)^2+\omega^4+2 \omega^2 \left(7 \delta_\nu ^2-\bar\omega^2_\nu\right)\right]}\,.
\end{equation}
Similarly, we find analytic expressions for the energy of the RC, $\lambda_\nu^2 = \delta_\nu^2+\bar\omega_\nu^2$, and the coupling strength 
$g_\nu^2 = \Gamma_\nu \delta_\nu^2 / (64 \lambda_\nu)$.

\section{\label{app:FullCounting} Full counting statistics and large deviaton theory}

We consider systems described by $H'_\text{tot} =  H_\text{S'} + H_\text I + \sum_\nu H_\text{B}^\nu$, 
where $ H_\text I = -\sum_{\nu,n} U_n^{\nu} \sqrt{\bar\varepsilon_n \lambda_\nu}(e_n^\dagger -e_n)P_\nu$ and $\bar H_\text B^\nu =  \sum_k\Omega_{k_\nu}d_{k_\nu}^\dagger d_{k_\nu}$ [see Eq.~(5)]. Let us introduce a generalized density matrix~\cite{RestrepoEtAlNJP2018}
\begin{equation}
\bar \varrho_\text{tot} \left(\left\{\chi_n\right\}, t\right) \equiv \bar U\left(\left\{\chi_n\right\}, t\right)\bar \varrho_\text{tot}(0) \bar U^\dagger\left(\left\{\chi_n\right\},t\right),
\end{equation}
with factorizing initial density matrix $\bar \varrho_\text{tot}(0)=\bar \varrho(0) \otimes \sum_\nu \bar \varrho_\text B^\nu$, where $\bar \varrho_\text B^\nu \sim e^{-\beta^\nu H_\text B^\nu}$. Here, we have introduced the so called counting fields $\chi_n$ corresponding to the transport channel $n$. The modified evolution operator $\bar U(\{\chi_n\},t)$ is related to the usual evolution operator $ U(t)$ corresponding to $ H_\text{tot}$ by
\begin{equation}
\bar U\left(\left\{\chi_n\right\},t\right) = \exp\left(-\frac{i}{2} H_\text B^\mu \sum\limits_n \chi_n\right)  U(t)\exp\left(\frac{i}{2} H_\text B^\mu \sum\limits_n \chi_n\right),
\end{equation}
where $\mu$ denotes the reference reservoir. The modified reduced density matrix $\bar \varrho(\{\chi_n\},t)=\text{Tr}_\text B\{\bar \varrho_\text{tot}(\{\chi_n\},t)\}$ evolves according to a generalized master equation~\cite{EspositoHarbolaMukamelRMP2009},
\begin{equation}
\label{eq:ChiLindbladEquation}
\begin{aligned}
\dot{\bar \varrho} =& \mathcal L\bar\varrho +\sum\limits_n \left[F_n^\mu \left(e^{\text i\chi_n\bar \varepsilon_n}-1\right)e_n\bar\varrho e_n^\dagger +  G_n^\mu \left(e^{-\text i\chi_n\bar \varepsilon_n}-1\right)e_n^\dagger\bar\varrho e_n\right], 
\end{aligned}
\end{equation}
which can be derived by performing the usual perturbative expansion up to second order in $ H_\text I$. Here, $F_n^\mu$ and $ G_n^\mu$ are defined as in Eq.~(6). Note that for $\chi_n = 0$, $\bar \varrho = \varrho$ and the standard Lindblad master equation is recovered.  

The moment generating function associated to the probability distribution $p(\Delta E)=p(E_t-E_0)$ of two projective measurements of $ H_\text B^\mu$ at time $0$ with outcome $E_0$ and at time $t$ with outcome $E_t$ is given by~\cite{EspositoHarbolaMukamelRMP2009, RestrepoEtAlNJP2018}
\begin{equation}
\begin{aligned}
\mathcal M_\mu\left(\left\{\chi_n\right\}, t\right) &=\text{Tr}\left\{\bar \varrho\left(\left\{\chi_n\right\},t\right)\right\}\\ 
&= \int d\Delta E~\prod\limits_n e^{-\text i\chi_n\Delta E}p\left(\Delta E\right) \\
&=\prod\limits_n \mathcal M_\mu\left(\chi_n,t\right), 
\end{aligned}
\end{equation}
where the last equality holds for weak coupling in the parallel oscillator picture, since all transport channels are uncoupled. Thus, the statistics of the exchanged energy between the reference reservoir $\mu$ and the harmonic oscillators are completely independent from each other. In the long time limit, large deviation theory applies [86-90] and the moment generating function tends to~\cite{EspositoHarbolaMukamelRMP2009}
\begin{equation}
\mathcal M_\mu(\chi_n,t) \to e^{t\mathcal C_\mu^\infty}
\end{equation}
with (scaled) cumulant generating function (CGF)
\begin{equation}
\begin{aligned}
\mathcal C_\mu^\infty = \lim\limits_{t\to \infty}\frac{\ln\left(\text{Tr}\left\{\bar \varrho\left(\left\{\chi_n\right\},t\right)\right\}\right)}{t}.
\end{aligned}
\end{equation}
When investigating transport statistics of non-equilibrium systems, cumulants usually grow linearly in time~\cite{VaradhanCPApplMAth1966, GartnerTPA1977, EllisAnnProb1984, GarrahanLesanovskyPRL2010, PigeonEtAlPRA2015} and it is more convenient to investigate $\mathcal C_\mu^\infty$, which is scaled by the time $t$ between the two projective measurements. For the model at hand, i.e. harmonic oscillators independently coupled to bosonic reservoirs (see main text), the CGF takes the form of Eq.~(7).

\end{document}